\documentclass[10pt,letterpaper]{article}
\usepackage[top=0.85in,left=2.75in,footskip=0.75in,marginparwidth=2in]{geometry}

\usepackage[utf8]{inputenc}

\usepackage{cite}

\usepackage{nameref,hyperref}

\usepackage[right]{lineno}

\usepackage{microtype}
\DisableLigatures[f]{encoding = *, family = * }

\raggedright
\usepackage{changepage}

\usepackage[aboveskip=1pt,labelfont=bf,labelsep=period,singlelinecheck=off]{caption}

\makeatletter
\renewcommand{\@biblabel}[1]{\quad#1.}
\makeatother

\usepackage{lastpage,fancyhdr,graphicx}
\usepackage{epstopdf}
\fancyheadoffset[L]{2.25in}
\fancyfootoffset[L]{2.25in}

\usepackage{color}

\definecolor{Gray}{gray}{.25}

\usepackage{graphicx}

\usepackage{sidecap}

\usepackage{wrapfig}
\usepackage[pscoord]{eso-pic}
\usepackage[fulladjust]{marginnote}
\reversemarginpar

\begin{document}
\vspace*{0.35in}

% title goes here:
\begin{flushleft}
{\Large
\textbf\newline{Human-centered trust framework: An HCI perspective}
}
\newline
% authors go here:
\\

Sonia Sousa\textsuperscript{1, *2},
Jos\'e Cravino\textsuperscript{3},
Paulo Martins\textsuperscript{2},
David Lamas\textsuperscript{1}
\\
\bigskip
\bf{1} School of digital technologies, Tallinn University, Tallinn, Estonia 
\\
\bf{2} Universidade de Tr\'as-os-Montes e Alto Douro \& INESC TEC
\\
\bf{3} Universidade de Tr\'as-os-Montes e Alto Douro \& CIDTFF
\\
\bigskip
* sonia.sousa@tlu.ee

\end{flushleft}

\section*{Abstract}
The rationale of this work is based on the current user trust discourse of Artificial Intelligence (AI). We aim to produce novel HCI approaches that use trust as a facilitator for the uptake (or appropriation) of current technologies. We propose a framework (HCTFrame) to guide non-experts to unlock the full potential of user trust in AI design. Results derived from a data triangulation of findings from three literature reviews demystify some misconceptions of user trust in computer science and AI discourse, and three case studies are conducted to assess the effectiveness of a psychometric scale in mapping potential users' trust breakdowns and concerns. This work primarily contributes to the fight against the tendency to design technical-centered vulnerable interactions, which can eventually lead to additional real and perceived breaches of trust. The proposed framework can be used to guide system designers on how to map and define user trust and the socioethical and organisational needs and characteristics of AI system design. It can also guide AI system designers on how to develop a prototype and operationalise a solution that meets user trust requirements. The article ends by providing some user research tools that can be employed  to measure users' trust intentions and behaviours towards a proposed solution.

% stop line numbers
%\nolinenumbers

% the * after section prevents numbering
\section*{Introduction}
The recent exponential growth in artificial intelligence (AI) applications across industries, businesses, organisations, and households has brought benefits but also raised concerns regarding potential threats, disinformation, and the biased predictions that AI solutions can bring to society. As the popularity of AI increases, its potential to create opaque practices will also increase, making it difficult for such applications to be audited, externally verified, or questioned. It will also be difficult to predict AI behaviours and identify errors that can lead to biased decisions, unforeseen risks, and further user trust concerns \cite{ Araujo:2020dn, lopes2022hci, hickok2021lessons, zhang2021manage}. 

When mixed with poor design practices, these characteristics of AI, i.e., opacity and unpredicatability, can result in vulnerable interactions that lead to further breaches of trust (both real and perceived). Thus, current increase discourse on Human-centered AI (HCAI) and trustworthy AI (TAI) seeks to establish clear boundaries in terms of user trust and socioethical characteristics to overcome potential AI threats and vulnerabilities and to complement already-existing data protection regulatory mechanisms (GDPR - ISO/IEC 27001) \footnote{https://gdpr-info.eu/} \cite{xu2021human, wired, hickok2021lessons, IBM, floridi2021ethical, sundar2020rise}. 

However, despite the increased attention given to trust in computing in the two past decades, the body of literature on trust in computer sciences presents divergent notions of trust, an oversupply of trust models framed toward a context-specific application, and ways to measure it that focus on initial trust inter-relationships \cite{Sousa-confio21, ajenaghughrure2020measuring}. This unclear notion of trust is also shown in current user trust discourse in relation to AI, where there is a need to clarify the role of user trust in AI within the framework of current socioethical considerations, technical and design features, and user characteristics \cite{tita-amna22, vianello2022improving}. Additionally, there is a need to better illustrate how trust and risk inter-relate in AI. This concept was pushed forward by the recent EU AI act \footnote{{\url https://artificialintelligenceact.com/}} and its risk-based approach (i.e., unacceptable risk, high risk, \& limited risk). Finally, the trust notions of new users should be studied to determine the antecedents to users’ cognitive and emotional trust across culture, physical, and digital borders \cite{glikson2020human}. Otherwise, the complexity of AI (e.g., its opacity and autonomy) mixed with unclear notions of user trust will make it challenging to address novel HCAI and TAI design mechanisms to ensure that AI systems are perceived by humans as reliable, safe, and trustworthy \cite{shneiderman2020bridging}. Potential misconceptions about how user trust in AI affects long-term relationships and acceptance could also arise \cite{shin2021effects}.
 
The above-described challenges associated with user trust in AI provide the rationale for this work. There is a need to develop a novel HCI lead framework to guide non-experts to unlock their full potential in terms of user trust in AI design and ensure uptake (or appropriation) by users without fear.  The goal of this work was therefore to illustrate how the proposed Human-Centered Trustworthy Systems framework (HCTFrame) supports the trust design (RQ). 

We believe that the recent EU AI act will increase the need to consider the influence of user trust in AI design. In addition, this framework will help to clarify the inter-relations between user trust and socioethical considerations, technical and design features, and user characteristics. Ultimately, we hope that this framework can prevent past mistakes associated with forward-push data protection regulatory mechanisms (e.g., GDPR - ISO/IEC 27001)\footnote{https://gdpr-info.eu/} that did not fully consider the effects of technical implementation on future relationships. %Please check intended meaning is retained.
That led to potential future threats being missed. For example, Facebook and \emph{Cambridge Analytics}\footnote{https://www.bbc.com/news/technology-54376327} used malevolent privacy techniques to persuade individuals to abdicate their privacy and ethical principles \cite{jouhki2016facebook, hansen2007marrying}. 

The overall aim of this framework is to provide AI providers, designers, and other stakeholders with easy-to-follow guidelines on how to account for user trust in AI design. The framework is derived from data triangulation of findings from three literature reviews on misconceptions of user trust in AI discourse. Three case studies are used to evaluate the effectiveness of using a psychometric scale to map potential technological trust breakdowns and concerns \cite{Sousa:2021mz, Sousa22JMIR, pinto2022trust, pinto2020adaptation, tita-amna22, Sousa-confio21, ajenaghughrure2020measuring}.

\section*{TAI Applications \& challenges}
The increased trust in AI discourse has led to various perspectives on how user trust or Human trust in AI can shape or calibrate the user--AI relationship. Solutions that use TAI to minimise incorrect decisions, bias predictions, warfare, surveillance malpractices, and disinformation have been sought. In addition, research has been done on ways to shape users' cognitive and emotional trust and foster accountability \cite{Davis2020, Paez:2019nm, lopes2022hci, Araujo:2020dn, hickok2021lessons, ashby2019fourth, zuboff2019surveillance, marcus2019rebooting, glikson2020human, rossi2018building}

However, current findings and strategies that address TAI do not clearly show how user trust is inter-related with AI applications. For example, in \cite{shin2021effects}, a discussion on the use of AI explanations to foster trust in users is presented. Additionally, this area of research is covered by \cite{Shneiderman:2020hv}. A related review \cite{jobin2019artificial} on AI ethics guidelines in different countries revealed similarities in some attributes (e.g., transparency, justice and fairness, non-maleficence, responsibility, and privacy) but differences in the interpretation, prioritisation, and implementation of ethical principles. Sometimes approaches that are complementary in nature are presented and trust is viewed as the culmination of indirect attributes such as privacy \& accountability, safety \& security, transparency \& explainability, fairness \& non-discrimination, human control of technology, professional responsibility, promotion of human values, and international human rights \cite{fjeld2020principled}.  

The widespread popularity of AI has created  tension between the current state of AI development, deployment, and user trust in AI \cite{Davis2020}. Evidence of this can be found in \cite{tita-amna22}, where it is indicated that there is a need for more clarity on how user trust in AI is related to socioethical considerations, technical and design features, and user characteristics. In addition, \cite{glikson2020human} revealed the important role of tangibility, transparency, reliability, and immediacy behaviors in AI, as these characteristics are associated with the development of cognitive trust. The anthropomorphism of AI specifically plays a role in emotional trust. In \cite{ajenaghughrure2020measuring} and \cite{glikson2020human}, it is highlighted that trust in computing science has been addressed from a  technical-central point of view, which has led to narrow solutions with very specific applications, and trust in long-term relationships has not been established. Research with a human-central or socio-ethical point of view has led to an unclear notion of the role of user trust in improving the trustworthiness of AI Solutions \cite{Shneiderman:2020hv, vorm2022integrating}.

In conclusion, following the approach of the Special Interest Group in 1982 to address the design of Human-Computer Interactions (e.g., SIGCHI), there is a need for the development of novel  Human-Computer Interaction design approaches to  tackle the problems described.

\section*{The HCTFrame framework} 
The above visions, challenges, and trends associated with user trust in AI illustrate the complexity of AI System design. They also illustrate why system design is no longer seen as a single technical element but rather as a complex set of components working together to facilitate a common sociotechnical purpose. In the structure of sociotechnical systems, AI system design should consider the Human element (people), the social and organisational structure, and the technical part of the system. Tasks should be recognized and prioritized according to their perceived sociotechnical importance and organisational and external changes \cite{abdelnour2019theorizing}. This vision argues, therefore, that coping with potential system failures does not just involve deviation from a specification (e.g., robust, reliable, secure, etc.) but rather judgment of the system's expectations and experiences (e.g., trustworthy (= able to be trusted), ethical, etc.).  

This information was obtained from several reviews that were undertaken to understand how trust and technology are addressed in HCI literature \cite{ajenaghughrure2020measuring, tita-amna22, Sousa-confio21}. Additionally, the influence of user trust on HCI interaction processes such as openness \cite{sousa11cpscom}, knowledge sharing \cite{sousa11ICWL}, privacy awareness \cite{Sousa13, oper2020}, collaboration \cite{Sousa14TEC}, cross-cultural factors \cite{Sousa:2021mz, Sousa:2021pm} was investigated. We also carried out several studies on the modelling of trust in AI-enabled systems \cite{Gulati19, ajenaghughrure2020risk, pinto2020adaptation}. 

\begin{figure}[h]
\centering
\includegraphics[width=1\linewidth]{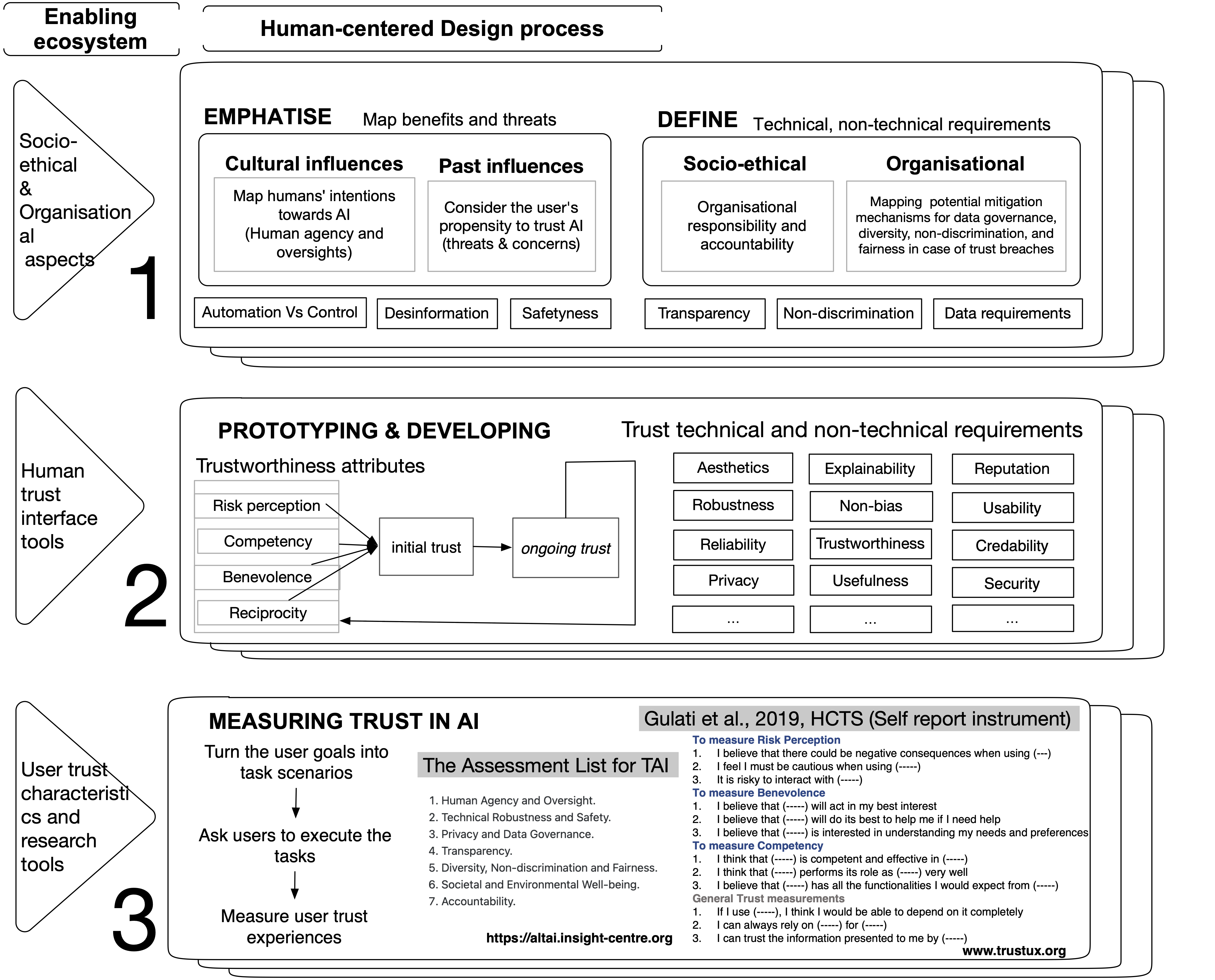} 
\caption{The interconnections among the four layers of the HCTframe analysis.} \label{All-HCTFrame},
\end{figure}

 The proposed framework (see figure \ref{All-HCTFrame}) has three layers of analysis. First, the socioethical and organisational characteristics of the AI system design are addressed. Then, the human element (people) is considered through the use of a prototype and a solution is developed to meet user trust requirements. Finally,  the system is judged by assessing the trustworthiness of its AI qualities.

\begin{figure}[ht]
\centering
\includegraphics[width=.8\linewidth]{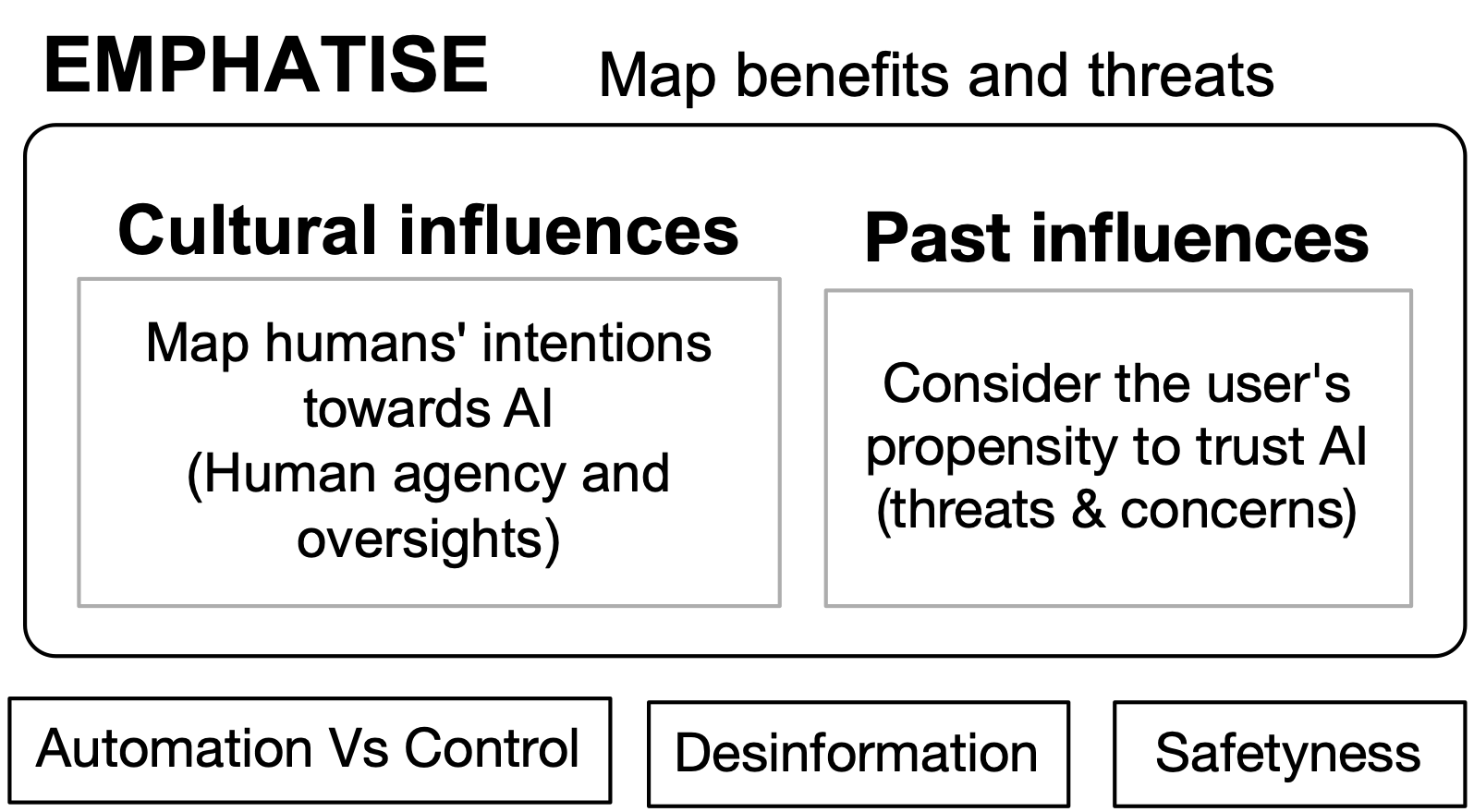} 
\caption{The interconnection between user's trust in AI and current socio-ethical and organisational requirements} \label{ph1-HCTFrame},
\end{figure}

{\bf Socioethical and organizational} layer 1: This is used in part 1 of the analysis to establish the connections between current regulations and trustworthy AI principles and users' socioethical needs based on principles like transparency, accountability, non-discrimination, and fairness. Here, the main goal is to establish links between users' sociotrust needs, AI characteristics, and existing regulatory principles. This broader view should illustrate users' trust perceptions across cultures and their implications (needs and threats) within society.  Potential threats and benefits are mapped for current and future generations (e.g., Human agency and oversights). Possible outcomes can be represented using affinity maps, personas, scenarios, or journey maps. Figure \ref{ph1-HCTFrame} illustrates the performance of this step. 

{\bf  Organisational trust requirements} layer 1: This is used in part 2 of the analysis to examine trust within an organisational structure. It focus on norms and roles in terms of how to act. The aim is to define the influence of trust within the corporate setting. Here, the main goal is to establish a link between regulatory principles and the organisational requirements that are necessary to ensure organisational responsibility requirements of the AI provider are satisfied. It should also help AI designers to rethink how user trust needs in AI intercept with organisational requirements. It involves the mapping of potential mitigation mechanisms for data governance, diversity, non-discrimination, and fairness in case of trust breaches. Possible outcomes can be represented in a document listing all of the technical and non-technical requirements needed to ensure a system's accountability. Figure \ref{ph1-HCTFrame} illustrates the performance of this step.

\begin{figure}[ht]
\centering
\includegraphics[width=.8\linewidth]{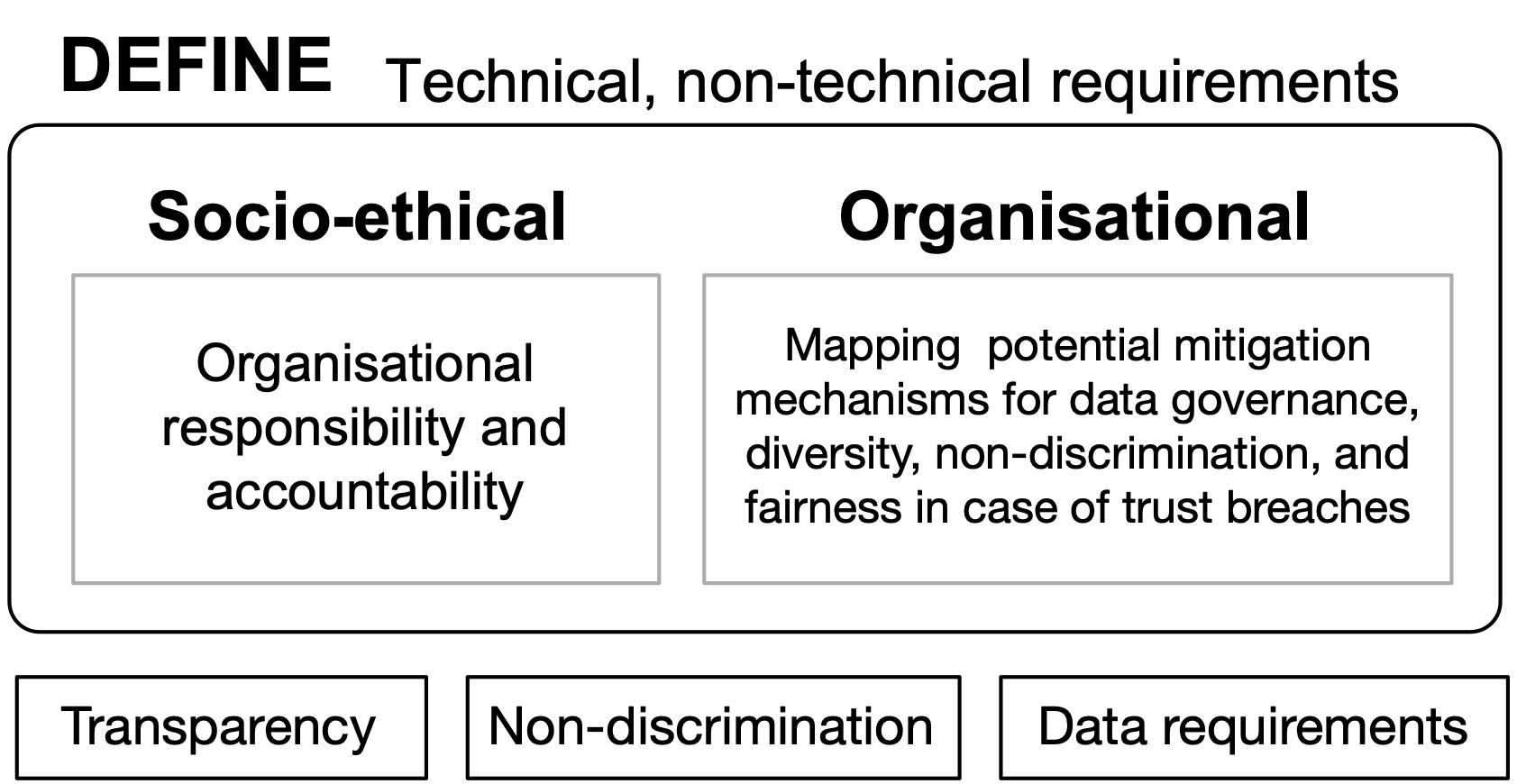} 
\caption{The interconnection between organisational responsibility and accountability for system outcomes} \label{ph2-HCTFrame}
\end{figure}

% Change to better clarify what was meant by Trust value Vs designed solution

\begin{figure}[ht]
\centering
\includegraphics[width=.8\linewidth]{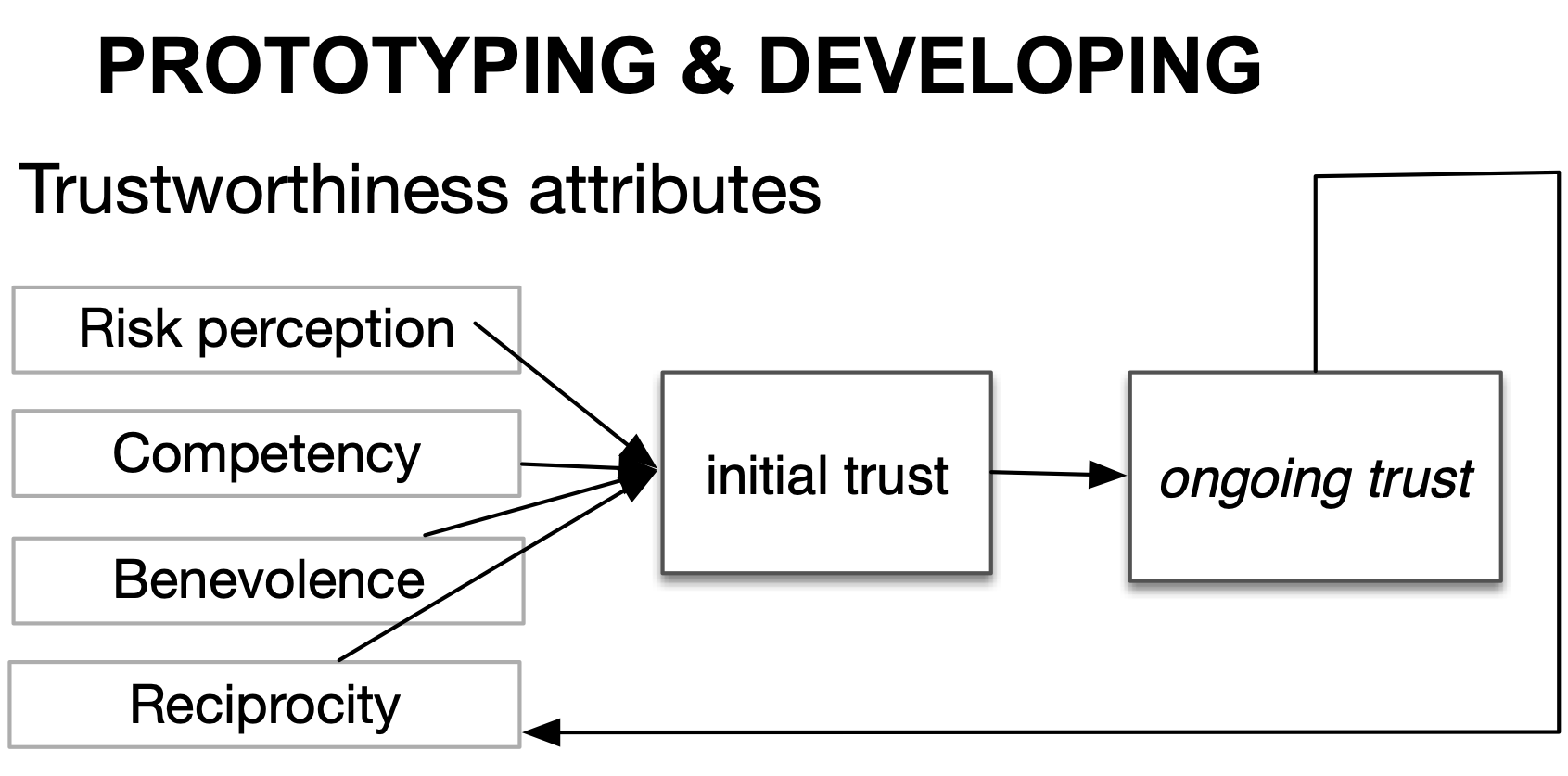} 
\includegraphics[width=1\linewidth]{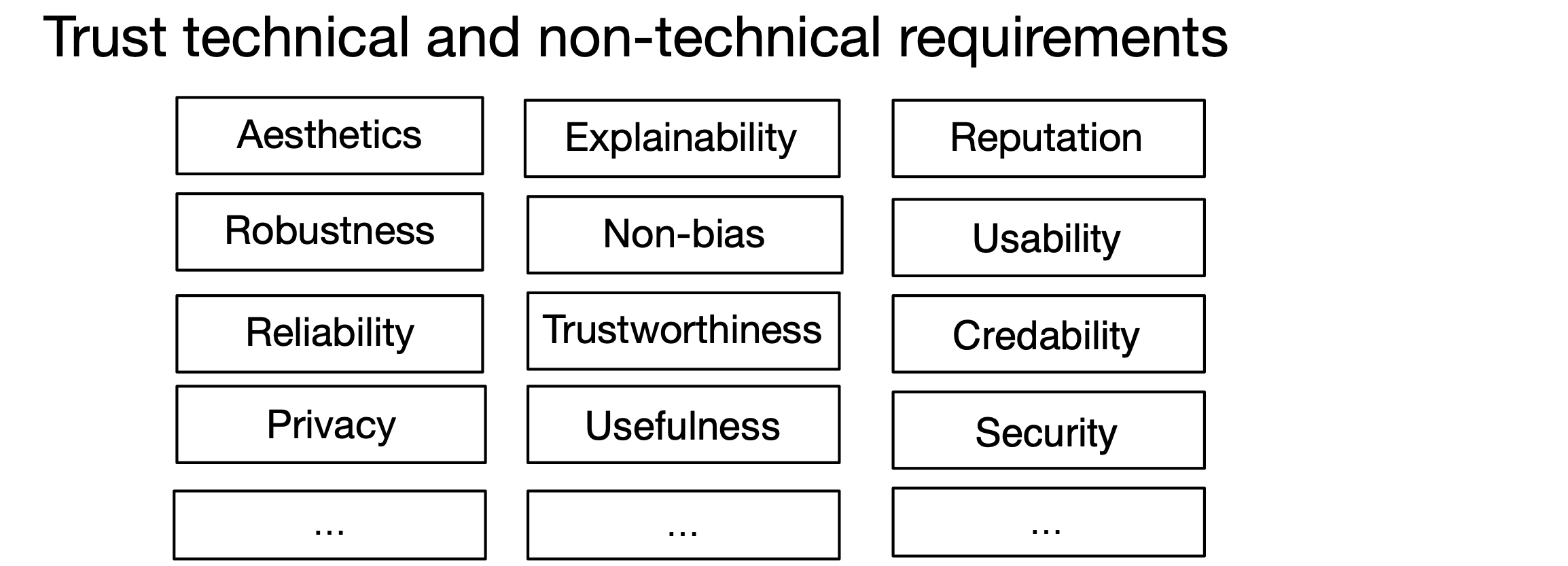} 
\caption{The interconnections between users' trust values and system-trust enabling solutions.} \label{ph3-HCTFrame}
\end{figure}

{\bf The Human trust element in the interface design}: This is include in layer 2 of the analysis. It combines user needs and system requirements into a set of interaction design space dimensions. The goal is to visualise the human (people) dimensions of the system and identify the user interface qualities of a system than can help to explain/describe the system's AI trust requirements.  This supports users in visualising system trust reassurance features that can assure users that the use of AI provides a system with minimised risks that is competent and benevolent. This follows the Human--Computer Trust model presented by \cite{Gulati19}. Possible outcomes include a set of design space features, reflected through a set of user stories and use cases. Figure \ref{ph3-HCTFrame} illustrates the performance of this step.

% Change to better clarify what was meant by Trust value Vs designed solution

{\bf Measuring user trust in AI}: This is included in layer 3 of the analysis. The goal is to determine whether the overall human--computer trust system design strategy does in fact foster user trust in an AI solution. Additionally, it can provide diagnoses or uncover possible AI trust design issues. This is carried out by implementing a user experience evaluation procedure to measure users' trust perceptions regarding the AI solution. The recommended procedure is to turn the user goals into task scenarios that enable the user to understand the solution's actions and functions and can explain the context behind  the need for trust. Then, a self-report instrument is provided to measure user trust in the system (see the Human--Computer Trust scale presented in \cite{Gulati19}). This instrument can be complemented with interviews and expert reviews using the Assessment List for Trustworthy AI (ALTAI) \footnote{https://altai.insight-centre.org/}. Figure \ref{ph4-HCTFrame} illustrates the performance of this step.  Possible outcomes can include a trust severity ratings associated with TAI problems. Those severity issues can refer to the frequency with which a problem can occur, its impact, and recommendations on how to fix it.

% Change and merge it in 

\begin{figure}[ht]
\centering
\includegraphics[width=.5\linewidth]{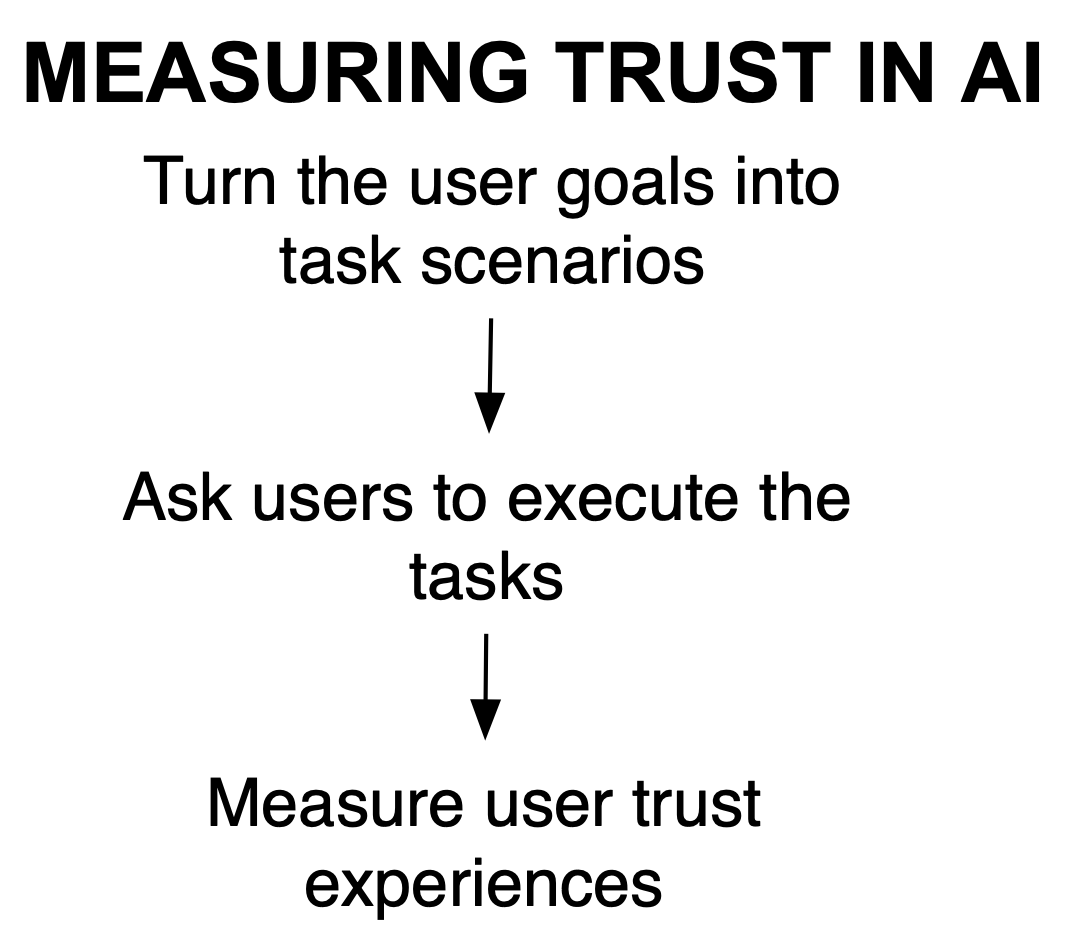} 
\caption{Measuring user's trust in AI} \label{ph4-HCTFrame}
\end{figure}

\begin{figure}[ht]
\centering
\includegraphics[width=.4\linewidth]{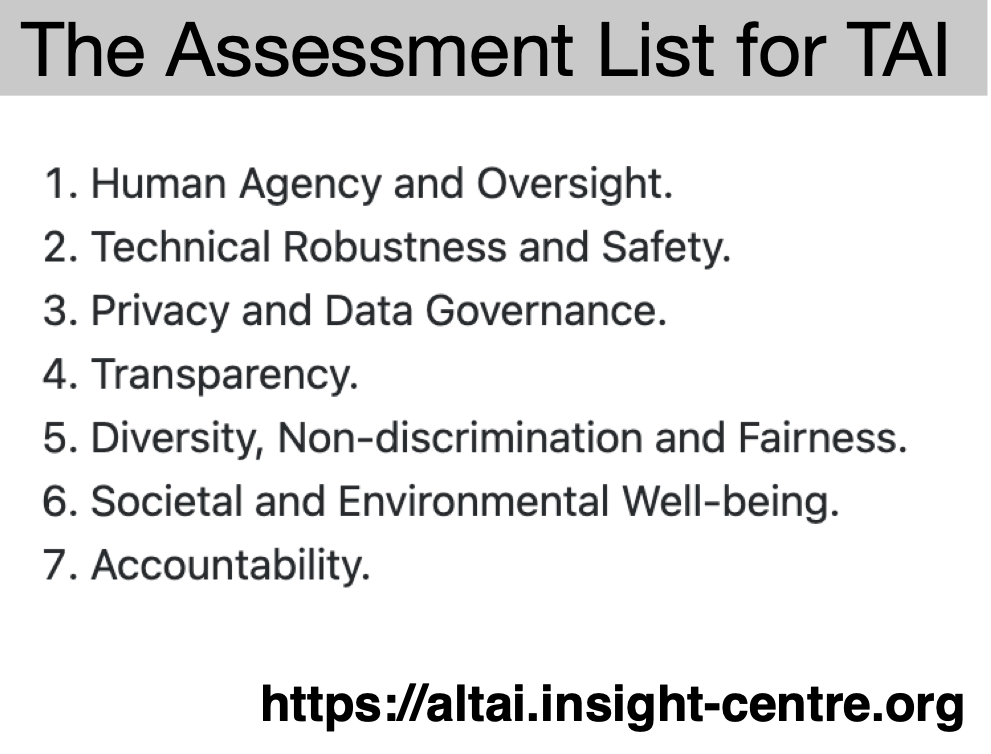} 
\includegraphics[width=.4\linewidth]{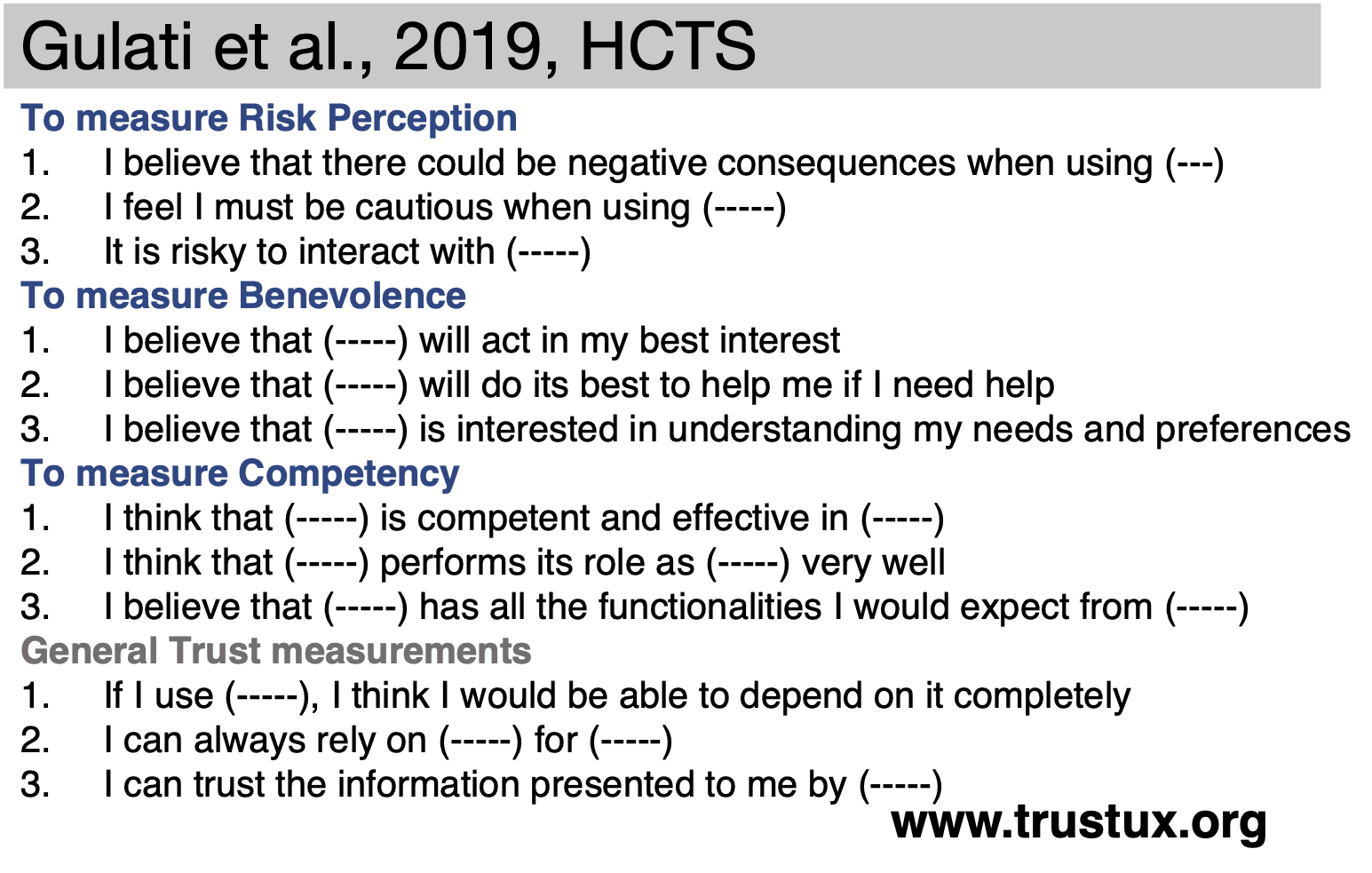} 
\caption{Measuring user's trust in AI} \label{ph4-HCTFrame}
\end{figure}

The following figure \ref{all-HCTframe} illustrates how the different layers of analysis complement each other.

\begin{figure}[ht]
\centering
\includegraphics[width=1\linewidth]{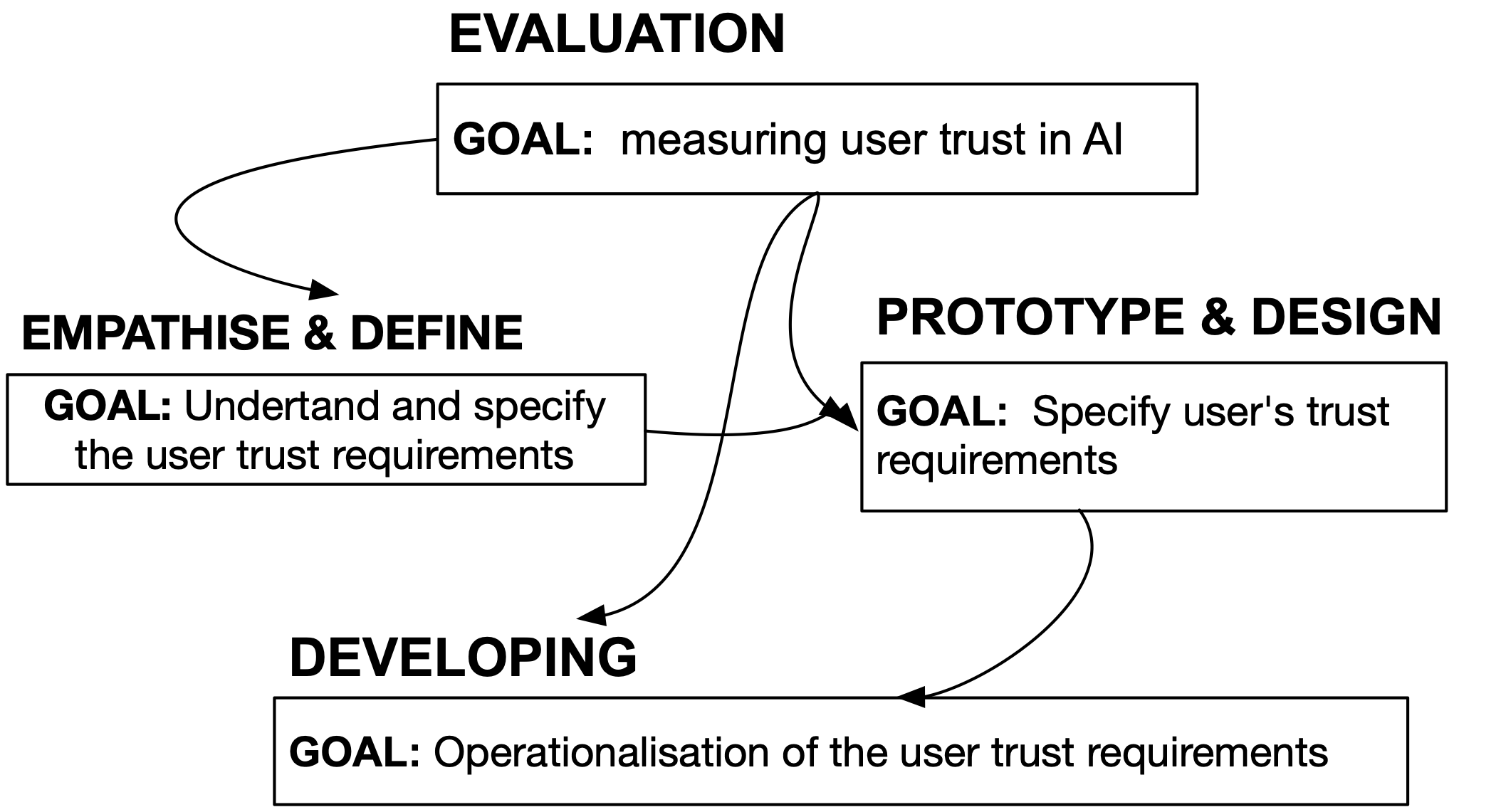} 
\caption{The interconnections among the different layers of analysis.} \label{all-HCTframe}
\end{figure}

\section*{The HCTframe design space}
As addressed above, recent efforts to address user trust in AI challenges, particularly the crescent discourse on TAI, guided this work. The above-proposed rationale was built from the need for novel HCI theories and frameworks to  guide non-experts to unlock their full potential in terms of user trust in AI design and ensure the uptake (or appropriation) of users without fear. This follows the recent increase in AI regulations and efforts to align trustworthy AI with Human-Centered design (HCD) approaches. This is demonstrated by the international standards for Artificial Intelligence (ISO/IEC JTC 1/SC 42/WG 3) \footnote{https://www.iso.org/committee/6794475.html} and initiatives such as those by the International Business Machines Corporation (IBM), the OECD (Organisation for Economic Co-operation and Development), and the European High-Level Expert Group on Artificial Intelligence (AI-HLEG) \footnote{https://digital-strategy.ec.europa.eu/en/policies/expert-group-ai}\cite{EUguidelines},  
as well as the EU AI act\footnote{https://digital-strategy.ec.europa.eu/en/policies/european-approach-artificial-intelligence}, the European Union's approach to boosting excellence and trust in AI. 

To overcome the potential threat of technology misuse, which can result in harm, we propose the use of HCTframe to guide non-trust experts in the transfer of HCI insights that focus on user trust in complex-enabled systems. The above-proposed framework aims to facilitate the incorporation of human trust into the TAI design process from an HCI perspective. The challenge is, however, determining how to frame multi-facet sociotechnical AI strategies within HCD and AI-enabled systems. This is particularly challenging when we consider that technology specialists have a lack of trust in current computer science education in terms of understanding how TAI practices can be addressed from a broad human-centered design point of view.

\begin{figure}[ht]
\centering
\includegraphics[width=.8\linewidth]{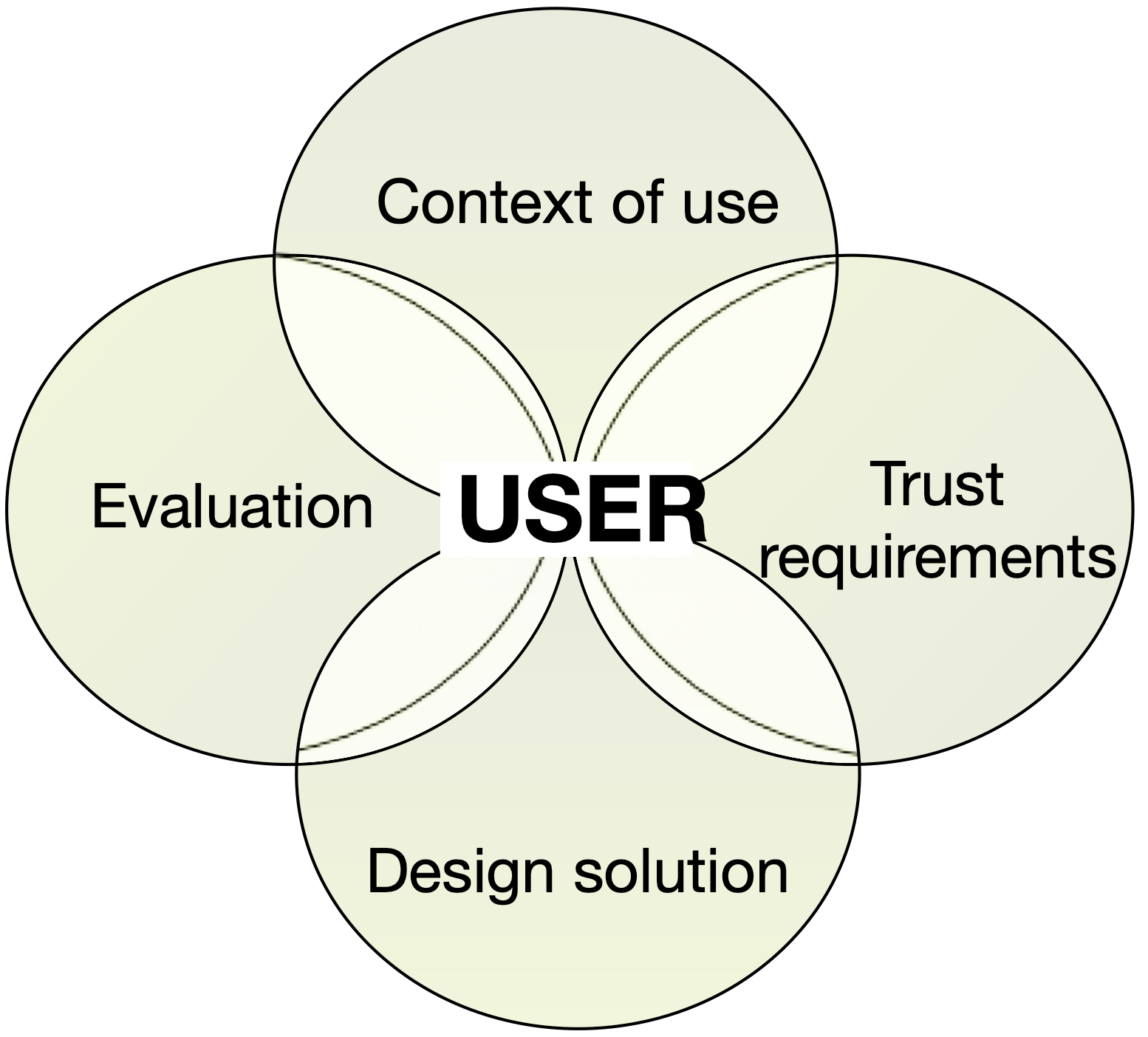} 
\caption{The interconnections among the different HCD design space components.} \label{HCD-process},
\end{figure}

ISO 9241-210\footnote{ISO 9241-210} presents the human-centered design (HCD) norms, besides understanding and specifying the context of use (e.g. system goals, stakeholders, user's characteristics, task, and environment).  This process requires the user requirements (e.g. context of use, the user needs, relevant user, interface knowledge, standards and guidelines) to be specified. Then, the design solutions, including user interactions and an interface, are produced. Finally, the design is evaluated against the requirements, including formative and summative evaluations such as a usability evaluation (efficiency, effectiveness and satisfaction criteria). Figure \ref{HCD-process} illustrates how the different HCD design space components intercept each other.

Therefore, the goal of the proposed HCTframe is to help non-experts to transfer user trust insights in the AI design process, as illustrated by figure \ref{HCD-process}. This can guide non-experts to address human trust in AI design and help to uncover root trust assumptions associated with AI technology. 

\subsection*{Validation, refinement, and reflection}
The goal of this research was to assess the extent to which the proposed HCTFrame can support HCI designers in mapping potential user trust needs, threats, and benefits. A design solution was proposed to meet user trust requirements. Finally, this was used to assess whether AI solutions were perceived as trustworthy by their users. The challenge was to use the framework to obtain actionable insights on three Cryptocurrency platforms: Coinbase\footnote{https://www.coinbase.com/}, Binance\footnote{https://accounts.binance.com/}, and tickeron\footnote{https://tickeron.com/}. In total, 25 participants were introduced to the AI design challenge and asked to use the framework to obtain actionable insights on user trust intentions and propose design recommendations.  A total of seven studies exploring the human--computer trust framework were conducted. Three groups focused on the Binance Cryptocurrency platform (11 participants), one on NFTs (3 participants), one on tickeron (4 participants), one on Coinbase (4 participants) and another on general  intentions and behaviours associated with investment in Cryptocurrency among Islamic female users (3 participants). 

This process was facilitated through a participatory design workshop implemented in October 2022. Participants' backgrounds ranged from computer science to psychology, design, and business. Participants were from different countries (e.g., Iran, Pakistan, Estonia, Nigeria, India, and Germany). Their aged ranged from 18 to 45 years old. Most participants were students from the Huma-Computer Interaction (HCI) Master program at Tallinn University. Overall, the feedback from participants was positive, and participants agreed that the  Human-Centered Trustworthy (HCTFrame) was able to guide them on how to map and define user trust and socioethical and organisational needs and characteristics related to AI system design.  This indicates that the framework is a useful tool that can facilitate trust design (RQ). This work also has some limitations. For example, participants could not link the technical and non-technical requirements of the platform with regulatory and organisational principles. Instead, their focus was on addressing Layers 2 and 3 of the framework and using task scenarios and the self-reported Human--Computer Trust scale (HCTS) as well as interviews to measure users' trust in cryptocurrency applications. There is a need for clear information on how the platform handles privacy and  digital threats and more guidance on the potential risks of investing in cryptocurrency for novice users.
Regarding the study of cultural differences, the findings indicate that some Islamic Women do not feel comfortable with investing in cryptocurrency,  despite shopping online regularly.  Other participants mentioned the need for clear information regarding the accountability and responsibility of the AI provider (Binance) in situations where something goes wrong. The majority of participants did not mention users' trust needs regarding data governance, diversity, and discrimination.  Additionally, the recommendations and insights provided were not detailed and cannot be generalised as the sample covered a specific population. In conclusion, the HCTFrame can provide insights into users' trust in AI systems, but it needs to be used in conjunction with other design tools.

\subsection*{The nature of HCI trust research}

The related body of literature shows that after two decades of investment and advances in trust concepts used in Computer science, we now place more importance on trust and trustworthiness notions, and the discourse has changed towards a human-centered point of view with a need to develop, deploy, and measure the quality of trust perceptions from an HCD perspective \cite{bauer2019conceptualizing}. 

Regarding the limitations of current evidence on trust measures, there has been an over-reliance on addressing user trust effects in terms of the initial intention to adopt AI systems (i.e., in the short term) with less exploration of the effects on long-term behavior (i.e., ongoing trust) \cite{glikson2020human, yang2022user}.  There is also a lack of clarity on how the measurement of user trust in AI-enabled systems can lead to distinct interpretations of inherent user trust characteristics (i.e., personality traits, gender, and self-trust) when addressing the user trust aspect in AI-enabling systems \cite{tita-amna22}.

This lack of clarity on trust notions and measures  makes this area of research susceptible to varied interpretations. At present, the different tools available to measure user trust can overwhelm and challenge non-experts when they need to evaluate their design solutions. A validated tool, such as the proposed HCTframe, i.e., the  Human-Computer trust scale (HCTS) may address this concern \cite{tita-amna22, Gulati19}.

The same applies regarding current practice in terms of addressing user trust over time \cite{yang2022user}. Qualitative methods are the most commonly used method to measure user trust in AI from a human-centered point of view \cite{barda2020qualitative, klumpp2019logistics}.  Although risk is a key factor that influences trust during the initial moment, by trusting, we assume that there is a potential gain, while by distrusting, we avoid a possible loss \cite{Bacharach2007}. 

Trust in this phase represents the trust assumptions or beliefs regarding the likelihood that another entity's future actions will be beneficial and favourable to the use \cite{Mcknight, luhmann2000familiarity}. In this phase, the main factors that play roles in the decision are the desire to accomplish a goal and motivation or willingness to engage in new situations (e.g., propensity for risk, openness attitudes, etc.). Initial trust is also known as swift trust \cite{robert2009individual}.

Trust should be viewed as a reassurance mechanism that supports users' intentions or predisposition to act and cooperate. Trust occurs following more than one encounter with the trustee after both parties have established a trust bond. Knowledge-based trust means that the trustor knows the other party well enough to predict the trustee's behaviour in a situation \cite{luhmann2000familiarity}.  Here, a key factor of trust influence is the context \cite{tita-amna22}.  This highlights the importance of selecting and tailoring features of the system according to the targeted user group’s characteristics and attributes, as addressed in the HCTframe. 

For example, it is important to determine which technical and design features should be emphasised according to the contexts and goals of the system's tasks \cite{tita-amna22}. This has led to debate on how intelligent technologies could become part of users' trusted experiences. It is important to ensure that AI characteristics are trustworthy and that users can maintain a bond over time. Current fears include AI failure and biased decision-making as potential threats, while the proposed framework of trustworthy AI principles can act as a design reassurance mechanism \cite{sousa2016value, Mcknight, Lankton:2015xf}. 

% Change tranform the checklist into text 

Nonetheless, the design of strategies to demystify AI adoption and to guarantee that humans do not misinterpret the causality of complex systems with which they interact \cite{Davis2020} has led to an oversupply of design toolkits, guidelines, checklists, and frameworks. Such tools aim to bring more clarity to this challenge but do not provide clarity on the user trust effect on AI. Design toolkits include IDEO, established by the entitled trust Catalyst Fund \footnote{https://www.ideo.com/post/ai-ethics-collaborative-activities-for-designers}, and the IBM Trustworthy AI, a human-centered approach\footnote{https://www.ibm.com/watson/trustworthy-ai}.%Please check intended meaning is retained.

Similarly, regarding TAI frameworks, a trustworthy AI was designed by \cite{smith2019designing}, an agile framework for producing trustworthy AI was detailed in \cite{leijnen2020agile}, and an article entitled Human-centered artificial intelligence: Reliable, safe \& trustworthy was presented by Shneiderman (2020) \cite{Shneiderman:2020hv}.  %Please check intended meaning is retained.
Despite recognizing the validity of existing mechanisms, the lack of clarity on how to address user trust in AI can mislead non-experts, meaning that they do not fully understand the intricate complexity of the subject and assume that trust is just a result of addressing indirect AI technical-centered characteristics, like reliability, or ensuring characteristics such as safeness, explainability, and reputation are present.  

The present work aimed to look beyond the AI view and rethink the effect of user trust on the intention to adopt AI.  Ultimately, we aimed to avoid the past mistakes associated with forward push privacy and data protection regulations (i.e,. GDPR) without considering the tension associated with the current state of technical innovations, profit-oriented deployment, user characteristics, and sociotechnical implications across societies.  As argued by \cite{rossi2018building} (p.132), to fully gauge the potential benefits of AI, trust needs to be established, both in the technology itself and in those who produce it. To put such principles to work, we need robust implementation mechanisms. 

\section*{Discussion}

Trustworthiness is complementary to trust. Trustworthiness reflects an individual's assessment of trust based on information provided. It is influenced by cultural assumptions and experiences, trust qualities that can evolve over time \cite{bauer2019conceptualizing}.

Needs and challenges that were not recognized in computer science trust literature 10 to 20 years ago are now a reality. This gap in the understanding of how to measure and address user trust characteristics over time can mean that IT and AI specialists do not clearly understand the complexity and application challenges associated with TAI system design as they are not equipped with the necessary skills to fully grasp the effect of trust on AI employment and the inherent risk associated with vulnerable design interactions that can lead to other breaches of trust, both real and perceived. 

Trust, on the other hand, is a personal quality, not a system quality, between a trustee (the entity being trusted) and the trustor (the person who trusts). Trusting involves a will to be vulnerable. Note that vulnerability implies risk acceptance based on how well the 'trustor' perceives the 'trustee' to be trustworthy \cite{mayer2006integrative}. According to \cite{cavoukian2012privacy}, privacy or the disclosure of certain types of information can help to foster trust in a system but should not be the main or only way to ensure a system's trustworthiness. Moreover, risk is a constant factor in the application context that can  significantly affect users' trust \cite{lee2004trust, heylighen2002complexity}. 

Trust is associated with the contemplation of subtle decisions and trust is associated with a time anchor that can be weakened or strengthened over time \cite{Bacharach2007, luhmann2000familiarity}. Therefore, trusting exists as a reinsurance element, i.e., a user's ability to identify whether an object or individual is trustworthy. Humans' ability to assess the risks or gains associated with being willing to engage in an interaction is what fosters trustworthiness in technology \cite{sousa2016value}. The ability to determine the trustworthiness of an object (e.g., an AI-enabling system) is affected by how individuals perceive the uncertainty of the situation (Risk assessment).  Some individuals are more willing than others to take risks \cite{kahneman2013prospect} describes. Similarly, individuals have different needs in terms of predicting the trustee's competency (i.e., ability to deliver the expected results) and assessing the intentions of others (i.e., benevolence).

Most common AI trust conceptualisations either follow the definitions of Mayer et al. (2006) \cite{Mayer95, mayer2006integrative}, 'the willingness of a party to be vulnerable to the actions of another party based on the expectation that the other will perform a particular action important to the trustor,
irrespective of the ability to monitor or control that
other party', or Lee and See (2004) \cite{lee2004trust}, 'the attitude that an agent will help achieve an individual’s goals in a situation characterized by uncertainty and vulnerability .

Thus, trusting involves a subtle decision based on the complexity of the game one finds themselves playing, and trust is associated with a time anchor that can be weakened or strengthened over time \cite{Bacharach2007, luhmann2000familiarity}. 

Thus, a trustful relationship requires the assessment of risks (i.e., gains and losses) over time. An evaluation of the tool's ability (e.g., competence, reliability, accuracy, etc.) to perform the desired outcomes and an assessment of the entity (i.e., company, organisation, institution, etc.) are required. Individuals expect that digital relationships will follow expected social norms and ethical principles (e.g., benevolence, reciprocity,  transparency, etc.).  The user characteristics that influence user trust (i.e., personality traits,
gender, and self-trust, etc.) should be mapped. User attitudes (i.e., user acceptance and readiness, needs and expectations, judgment and perceptions) should be assessed. External variables of the users (i.e., initial interactions, user interactions, cognitive load levels, time and usage) should also be assessed \cite{tita-amna22, Sousa-confio21}

\section*{Conclusion}

The above paragraph addresses current challenges and research gaps related to trust in technology and considers current computer science research gaps and challenges from an HCI perspective. We recognize that the complex nature of computing makes it challenging to incorporate an AI trust enabling ecosystem without the use of a transdisciplinary approach, which could elicit trust-breaking behaviours with potentially negative effects. We proposed the HCTframe with the aim of guiding non-trust experts to transfer HCI insights that focus on user trust into complex systems. We focused on the implementation and operationalisation of key trust requirements by examining users' trust influence from a broader point of view in a specific organisational setting. This involved the observation of users' social and trust "co-experiences" and interpretations of user needs by measuring how users perceive a system to be trustworthy.

We provided an overview of the nature of trust research in HCI and addressed trust inter-relationship in current complex systems. We hope to eventually support non-trust experts in their efforts to increase human trust in systems and supporting AI uptake (or appropriation) without fear.

We hope that this work will help clarify some contradictions between the current state of trust in computing research and the latest trustworthy AI applications. We also hope that our paper will give the reader a more tangible idea of the consequences of not addressing the human value.

%\clearpage

\section*{Acknowledgments}
This study was partly funded by the Trust and Influence programme under grant agreement number 21IOE051, the European Office of Aerospace Research and Development (EOARD), and the U.S. Air Force Office of Scientific Research (AFOSR).

\nolinenumbers

%This is where your bibliography is generated. Make sure that your .bib file is actually called library.bib
\bibliography{library}

%This defines the bibliographies style. Search online for a list of available styles.
\bibliographystyle{abbrv}

\end{document}